# Solid-Lubrication Performance of $Ti_3C_2T_x$ - Effect of Tribo-Chemistry and Exfoliation

Andreas Rosenkranz[1*], Bo Wang[2], Dario Zambrano[1], Javier Marqués Henríquez[1], Jose Y. Aguilar-Hurtado[1], Edoardo Marquis[3], Paolo Restuccia[3], Brian C. Wyatt[4], M. Clelia Righi[3] and Babak Anasori[4,5,6]

[1] Department of Chemical Engineering, Biotechnology and Materials, FCFM, University of Chile, Santiago, 8370415, Chile

[2] Chair of Functional Materials, Department of Materials Science & Engineering, Saarland University, Saarbrücken, 66123, Germany

[3] Department of Physics and Astronomy "Augusto Righi", University of Bologna, Italy

[4] Department of Mechanical and Energy Engineering and Integrated Nanosystems Development Institute, Purdue School of Engineering and Technology, Indiana University -Purdue University Indianapolis, Indiana 46202, United States of America

[5] School of Materials Engineering, Purdue University, West Lafayette, IN 47907, United States of America

[6] School of Mechanical Engineering, Purdue University, West Lafayette, IN 47907, United States of America

[*] Corresponding author

Andreas Rosenkranz (arosenkranz@ing.uchile.cl); ORCID: orcid.org/0000-0002-9006-1127

**Abstract**

Multi-layer $Ti_3C_2T_x$ coatings have demonstrated an outstanding wear performance with excellent durability due to beneficial tribo-layers formed. However, the involved formation processes dependent on the tribological conditions and coating thickness are yet to be fully explored. Therefore, we spray-coated $Ti_3C_2T_x$ multi-layer particles onto stainless steel substrates to create coatings with two different thicknesses and tested their solid lubrication performance with different normal loads (100 and 200 mN) and sliding frequencies (1 and 2.4 Hz) using linear-reciprocating ball-on-disk tribometry. We demonstrate that MXenes' tribological performance depends on their initial state (delaminated few-layer vs. multi-layer particles), coating thickness, applied load and sliding frequency. Specifically, the best behavior is observed for thinner multi-layer coatings tested at the lower frequency. In contrast, coatings made of delaminated few-layer MXene are not as effective as their multi-layer counterparts. Our high-resolution interface characterization by transmission electron microscopy revealed unambiguous differences regarding the uniformity and chemistry of the formed tribo-layers as well as the degree of tribo-induced MXenes' exfoliation. Atomistic insights into the exfoliation process and molecular dynamic simulations quantitatively backed up our experimental results regarding coating thickness and velocity dependency. This ultimately demonstrates that MXenes' tribological performance is governed by the underlying tribo-chemistry and their exfoliation ability during rubbing.

**1. Introduction**

Tribology, which involves friction, wear, and lubrication, directly connects to our daily life [1] and is a key discipline to tackle important challenges such as steadily rising $CO_2$ emissions and global warming [2, 3]. Considering diminishing raw oil resources and the dependency on oils and greases to reduce friction and wear, the tribological community is pushed to develop greener and more sustainable lubrication technology [2-4]. As a result, solid lubricant coatings based on 2D materials (graphene, graphene oxide, $MoS_2$, among others), which improve the frictional performance by either their low-shear strength and/or the formation of beneficial tribo-layers, have become a recent focus of tribological research [5-8]. However, these 2D materials still have material-specific limitations such as an environment-dependent frictional behavior (influence of humidity and/or vacuum) or a limited wear resistance due poor coating/substrate interfaces and/or weak secondary interactions in the out-of-plane direction based upon van-der-Waals interactions [6-9]. The latter makes them prone to wear-off once the first wear features are initiated, which affects their functionality and frictional performance. As a result, more work needs to be dedicated to the development of wear-resistant solid lubricant coatings.

In this context, early transition metal carbides, nitrides and carbonitrides (MXenes), a comparatively new family in the class of 2D materials discovered in 2011 [10-14], have shown promising friction and wear properties which can address the shortcomings of current state-of-the-art solid lubricants [5, 15-17]. MXenes are commonly abbreviated by the chemical formula $M_{n+1}X_nT_x$ ($n = 1$ to 4), which describes alternating layers of early transition metals (M: groups 3 to 6 of the periodic table) and layers of carbon/nitrogen (X). The outer transition metal layers have surface terminations, such as –O, –OH, –F, that are shown as $T_x$ in the formula [13]. In 2019, pioneering studies of Lian et al. [18] and Rosenkranz et al. [19] demonstrated the possibility to use multi-layer $Ti_3C_2T_x$ nano-sheets as solid lubricant coatings on copper and stainless steel substrates with a pronounced 4-fold friction reduction. Regarding the effect of tribological testing conditions, Marian et al. demonstrated that multi-layer $Ti_3C_2T_x$ coatings show an improved performance (50 % friction reduction) under low load (contact pressures up to 0.8 GPa) and low humidity (20 % relative humidity) [20].

In a later study, Grützmacher et al. demonstrated an outstanding tribological performance (stable and low COF of about 0.2 for 100.000 sliding cycles) for 100-nm-thick electro-sprayed coatings due to the formation of a beneficial MXene-derived tribo-layer [21]. After testing, this MXene-based tribo-layer contained structurally and chemically highly-degraded MXenes with random orientation, which were intermixed with amorphous and nanocrystalline oxidic species formed due to the cyclic testing. Further, the tribo-layer was initially formed on the substrate before being transferred to the tribological counter-body by adhesive processes. This ultimately changed the tribological system thus keeping friction low and stable for 100.000 sliding cycles representing an excellent wear lifetime, thus outperforming most of the existing state-of-the-art solid lubricants [21]. In another study comparing coatings fabricated with few- and multi-layer $Ti_3C_2T_x$ nano-sheets with the same coating thickness, Hurtado et al. demonstrated that only multi-layer MXene coating induce beneficial tribological effects [22]. While these initial results from Hurtado et al. are interesting, they did not present any in-depth analysis of the governing mechanisms and no further explanation for the obtained results.

Although this previous literature highlighting the tribological improvement of MXene-derived tribo-layers, little is known about the chemical composition and morphology of the formed tribo-layers as well as the influence of the operating conditions on their formation. Currently, existent literature demonstrates that exceeding a critical normal load (contact pressure) and high relative humidities are detrimental for the tribo-layer formation and result in increased friction and catastrophic wear [20]. However, the effect of the initial coating thickness and the adjusted sliding velocity/frequency is yet to be explored.

To explore the effect of these variables on the MXene-derived tribo-layer, we spray-coated multi-layer $Ti_3C_2T_x$ coatings with two different thicknesses onto stainless steel substrates and tested their solid lubrication performance at two normal loads (100 and 200 mN) and sliding frequencies (1 and 2.4 Hz) using linear-reciprocating ball-on-disk tribometry. To visualize the morphological changes in the MXene-derived tribo-layer, high-resolution transmission electron microscopy (TEM) was used to characterize the formed tribo-layers and interfaces. Based on the observed exfoliation tendencies, additional tribological measurements were performed on few-layer $Ti_3C_2T_x$ coatings with the same coating thickness with consistent tribological testing parameters. Finally, the experimental results were qualitatively compared with reactive molecular dynamics (MD) simulations to provide theoretical insights regarding the observed thickness- and velocity-dependency of the frictional response of the multi-layer $Ti_3C_2T_x$ coatings.

## 2. Experimental section

### *2.1 Synthesis and characterization of $Ti_3C_2T_x$ nano-sheets*

Multi-layer $Ti_3C_2T_x$ nano-sheets were made by selective etching of 2 g of MAX-$Ti_3AlC_2$ powder (purchased from Forsman Scientific Co. Ltd., Beijing, China) in 20 ml of highly-concentrated hydrofluoric acid (HF) with a concentration of 40 wt.-%. The mixture of $Ti_3AlC_2$ powder and HF was stirred at a speed of 60 rpm and a temperature of 35 °C for 24 hours. After etching, the mixture was centrifuged at 3500 rpm and washed with deionized water in several washing cycles until a final pH of 6 was obtained and the residue was collected. Finally, the suspension was vacuum-filtrated and freeze-dried for 24 hours at -60 °C (pressure below 30 Pa). To fabricate few-layer MXenes, centrifuged multi-layer nano-sheets were further exfoliated in distilled water by ultrasonication. Subsequently, the solution was centrifuged at 1550 rpm for 5 minutes before collecting the supernatant, which was processed and dried using the same protocol.

To characterize the as-synthesized multi-layer $Ti_3C_2T_x$ nano-sheets, TEM (Tecnai F20, FEI) was employed with an acceleration voltage of 200 kV. Surface-sensitive chemical characterization was performed by X-Ray photoelectron spectroscopy (XPS) using non-monochromatic $AlK_\alpha$ radiation at 15 kV and 400 W with a pass energy of 44.75 eV. Coarser scans between 1000 to 0 eV were realized with a step size of 1 eV, while more detailed scans of the $Ti_{2p}$, $O_{1s}$ and $C_{1s}$ peaks were performed with a higher resolution of 0.1 eV. Apart from the $Ti_{2p}$ peak, where linear background subtraction was applied, all peaks were corrected using the Shirley method.

### *2.2 Coating deposition*

For all few- or multi-layer $Ti_3C_2T_x$ nano-sheets samples, mirror-polished stainless-steel substrates (AISI 304) having a Young's modulus of 200 GPa, a hardness of 2 GPa, and a roughness ($R_q$) of about 120 nm were utilized. Prior to coating deposition, the freeze-dried MXene powders were dispersed in 20 ml ethanol in 2 and 5 mg/mL solutions. To improve the dispersion stability, the dispersions were shear-mixed using a homogenizer at 10.000 rpm for 5 minutes followed by ultrasonication for 3 hours. Afterwards, a volume of 2 ml was transferred to a self-built spray-coating setup, consisting of an airbrush pistol, a compressor, and a heating table. Spray-coating was performed with a nozzle-substrate distance of 15.5 cm and an air pressure of 0.15 N/mm$^2$. The substrate was pre-heated to 90°C to ensure fast solvent evaporation, which was necessary to prevent droplet formation to generate homogenous coatings. The thickness of the coatings was measured using a laser confocal microscope (Olympus LEXT 3D OLS400, 405 nm wavelength).

*2.3 Tribological experiments and tribo-layer characterization*

The solid lubrication performance of the fabricated $Ti_3C_2T_x$ coatings was evaluated using ball-on-disk tribometry (MFT 5000, Rtec Instruments) in linear-reciprocating sliding tests. The Hertzian contact pressure was varied between 0.3 and 0.5 GPa using two different normal loads of 100 and 200 mN. The stroke length was kept constant at 2.5 mm, while the sliding velocity was adjusted at 2.5 (1 Hz) and 6 mm/s (2.4 Hz), respectively. Temperature and humidity were kept constant at about 25 °C and 45 %, respectively. As tribological counterbody, a steel ball (AISI 52100, Rtec Instruments, hardness of 2.1 GPa and roughness Rq of about 470 nm) with a diameter of 4 mm was used. All experiments were repeated 5 times to calculate mean values and error bars.

After the tribological experiments, the resulting wear tracks/features created on the multi-layer coatings were first imaged by light microscopy (DIC, Zeiss Axioscope 5) using differential interference contrast (DIC) and white light interferometry (MFT 5000, Rtec Instruments). Based on the white light interferometric images, the respective wear volumes were estimated using the software Mountains Map. Cross-sectional transmission electron microscopy (TEM) samples of the tribo-layer were prepared *in-situ* of selected area from the wear tracks using focused ion beam microscopy (FIB, Verios G4 UC, Thermo Scientific, USA). TEM (Talos F200X, Thermo Fisher, USA). Further, we utilized energy dispersive spectroscopy (EDS) detector equipped in our TEM equipment to characterize the morphology and elemental distribution of the tribo-layer.

*2.4 Computational Modelling*

MD simulations were carried out using LAMMPS to model orthorhombic cells with the inclusion of periodic boundary conditions [23]. The basal-plane area was kept fixed to 36.39 Å x 31.51 Å, corresponding to a 12x12 supercell (720 atoms for each $Ti_2CO_2$ layer). The vertical height of the cell was varied from case to case to adjust the number of layers N included in the cell. The choice was done to guarantee a distance between replicas of at least 12 Å. All movable layers of the system have been thermalized at constant temperature (T = 300 K), except for the slider that was kept at a temperature of 1K to minimize atomic diffusion. The temperature of each layer was controlled during simulations by a Nosé-Hoover chain of thermostats. An integration time step of 0.1 fs was used, and each simulation run comprised 2.7 million MD steps. The friction force at time t was calculated as the mean value of the forces experienced by the slider's center of mass along the x direction, taking data every 10 ps from the beginning of the MD up to time t. This quantity is also known as cumulative average, as the mean value varies over the simulation time. After 100 ps, the cumulative average of the friction force starts to level off, while only minor oscillations were observed between 100 and 270 ps (please refer to **Figure S1** in the Supplementary Information for more details). The error bars are calculated as the difference between the maximum and minimum value of the running average from 100 ps onwards. The shear stress (in GPa) is calculated by dividing the friction force by the in-plane area of the supercell.

# 3. Results and Discussion

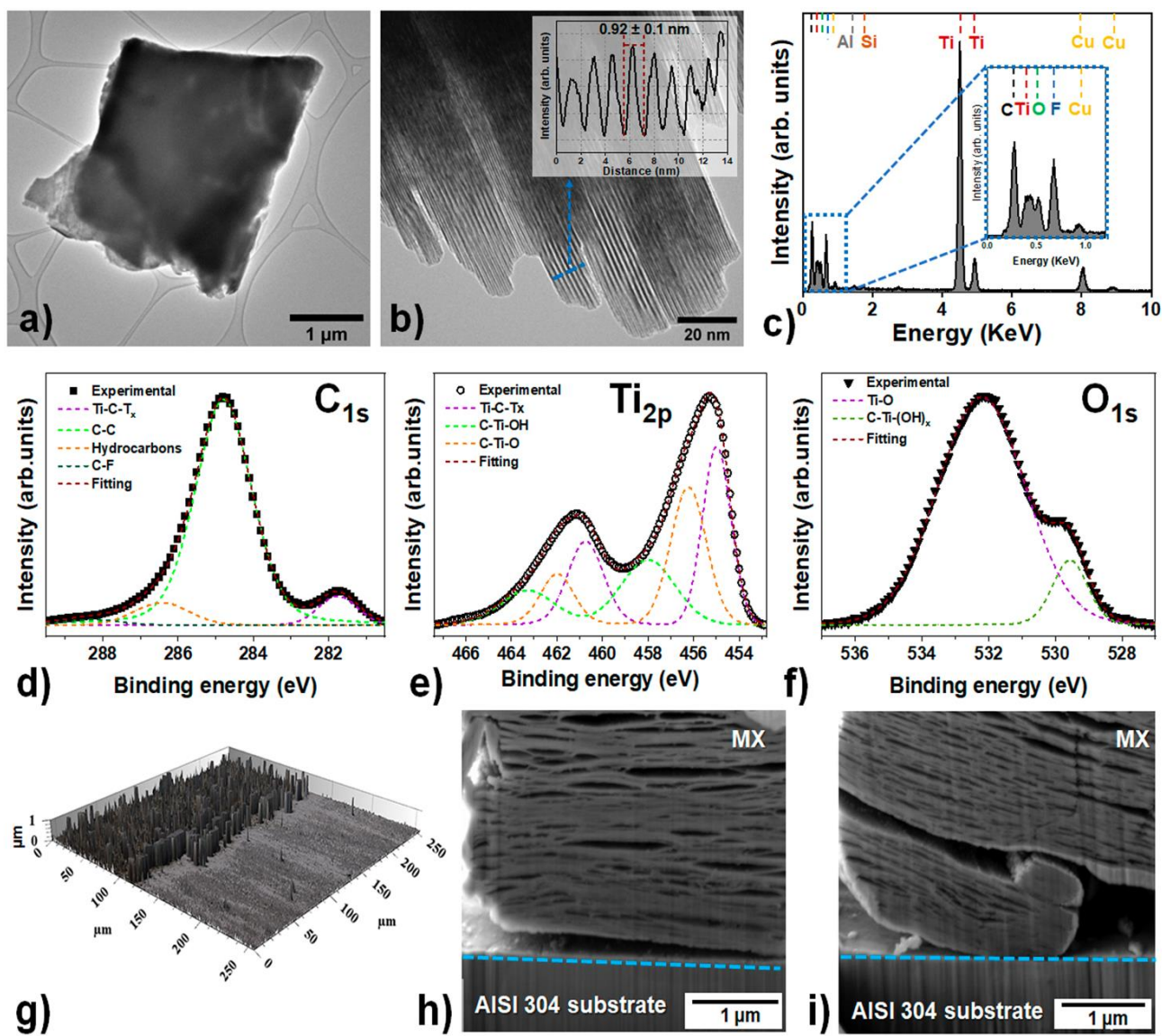

**Figure 1.** (**a**) Low- and (**b**) high-resolution TEM-micrographs of the as-synthesized multi-layer $Ti_3C_2T_x$ nano-sheets with (**c**) the corresponding elemental analysis conducted by TEM-EDX. Please note that the signals of Si and Cu in (**c**) originate from the detector and the sample holder, respectively. Spatially resolved and mathematically fitted X-ray photoelectron spectra of the (**d**) $C_{1s}$, (**e**) $Ti_{2p}$ and (**f**) $O_{1s}$ peak regions. Please note that different contributions necessary to fit the experimental data are shown in different colors with their corresponding assignment in the XP spectra. (**g**) Laser confocal microscopic image of the resulting multi-layer MXene coating (concentration of 5 mg/ml) deposited onto the stainless-steel substrate. (**h** and **i**) Cross-sectional field emission scanning electron micrographs of the initial orientation of the as-deposited nano-sheets with respect to the underlying substrate. Please note that multi-layer $Ti_3C_2T_x$ nano-sheets are abbreviated by "MX" in this figure.

The used HF etching of $Ti_3AlC_2$ resulted in multi-layer $Ti_3C_2T_x$ with x-y dimensions of about 3 microns (**Figure 1a**) and an interlamellar spacing in z-direction of 0.92 ± 0.1 nm (roughly the thickness of a flake of $Ti_3C_2T_x$ [24]) (**Figure 1b**). The elemental analysis by TEM-EDX (**Figure 1c**) confirmed titanium, carbon, oxygen, and fluorine as main elements, which align well with the overall stoichiometry and existing surface terminations (-O, -OH, and -F) [13, 14]. The presented TEM micrographs (**Figure 1a** and **b**) and only negligible traces of Al of 0.5 wt.-% (**Figure 1c**) likely from the parental $Ti_3AlC_2$ MAX-phase imply the successful synthesis of multi-layer $Ti_3C_2T_x$. In contrast, the as-synthesized few-layer $Ti_3C_2T_x$ nano-sheets had lateral x-y dimensions of about 1-1.5 microns and a highly reduced z-thickness as demonstrated in **Figure S2** (Supplementary Information). More detailed information about the characterization of the produced few-layer $Ti_3C_2T_x$ can be found in **[25-27]**.

To assess the superficial bonding states of the as-fabricated multi-layer $Ti_3C_2T_x$, spatially resolved XP spectra with the corresponding mathematical fits are shown for the $C_{1s}$ (**Figure 1d**), $Ti_{2p}$ (**Figure 1e**), and $O_{1s}$ (**Figure 1f**) peak regions. With respect to $C_{1s}$ (**Figure 1d**), the most dominant contribution centered at around 284.8 eV reflects C-C bonds and adventitious carbon [28], while the contribution centered at 281.8 eV connects with $C-Ti-T_x$ [28, 29]. Moreover, additional contributions relate to organic components (hydrocarbons) [29, 30] and C-F bonding states [28]. Related to $Ti_{2p}$ (**Figure 1e**), the most pronounced doublet (454.9 and 460.1 eV) can be connected with titanium carbides (MXenes) [31-33], while the other doublets represent C-Ti-OH [33] and C-Ti-O bonds [31, 33, 34], respectively. The fit functions needed to describe $O_{1s}$ (**Figures 1f**) can be assigned to metal oxides (Ti(IV) bonding states) and hydroxides ($C-Ti-(OH)_x$) [28, 29, 35].

After characterizing the as-synthesized nano-sheets, we spray-coated multi-layer $Ti_3C_2T_x$ coatings onto AISI 304 stainless-steel substrates (deposited area: 1 $cm^2$) using dispersion concentrations of 2 and 5 mg/ml, which resulted in average coating thicknesses of 400 ± 37 and 700 ± 65 nm, respectively. The overall coating quality and homogeneity were assessed by LCM (**Figure 1g**), which demonstrates the multi-layer MXene-coated substrate with the non-coated area. To elucidate the initial orientation of the multi-layer $Ti_3C_2T_x$ particles and the coating-substrate interface, transversal cross-sections (**Figure 1h** and **i**) were prepared by FIB and imaged by FE-SEM. These images show that the MXene sheets in the multi-layer particles are oriented mostly parallel to the substrate, in some cases very close to parallel (**Figure 1h**) and others, with an average inclination ranging between 15 - 20 degrees (**Figure 1i**). Irrespective of the initial orientation, the nano-sheets are only weakly adhered to the underlying substrate. This can be traced back to the spray-coating process, which does neither generate a preferential orientation of the nano-sheets deposited nor boosts the coating-substrate adhesion. Related to the coating fabrication using few-layer $Ti_3C_2T_x$, the same methodology was applied thus resulting in comparable coating thicknesses.

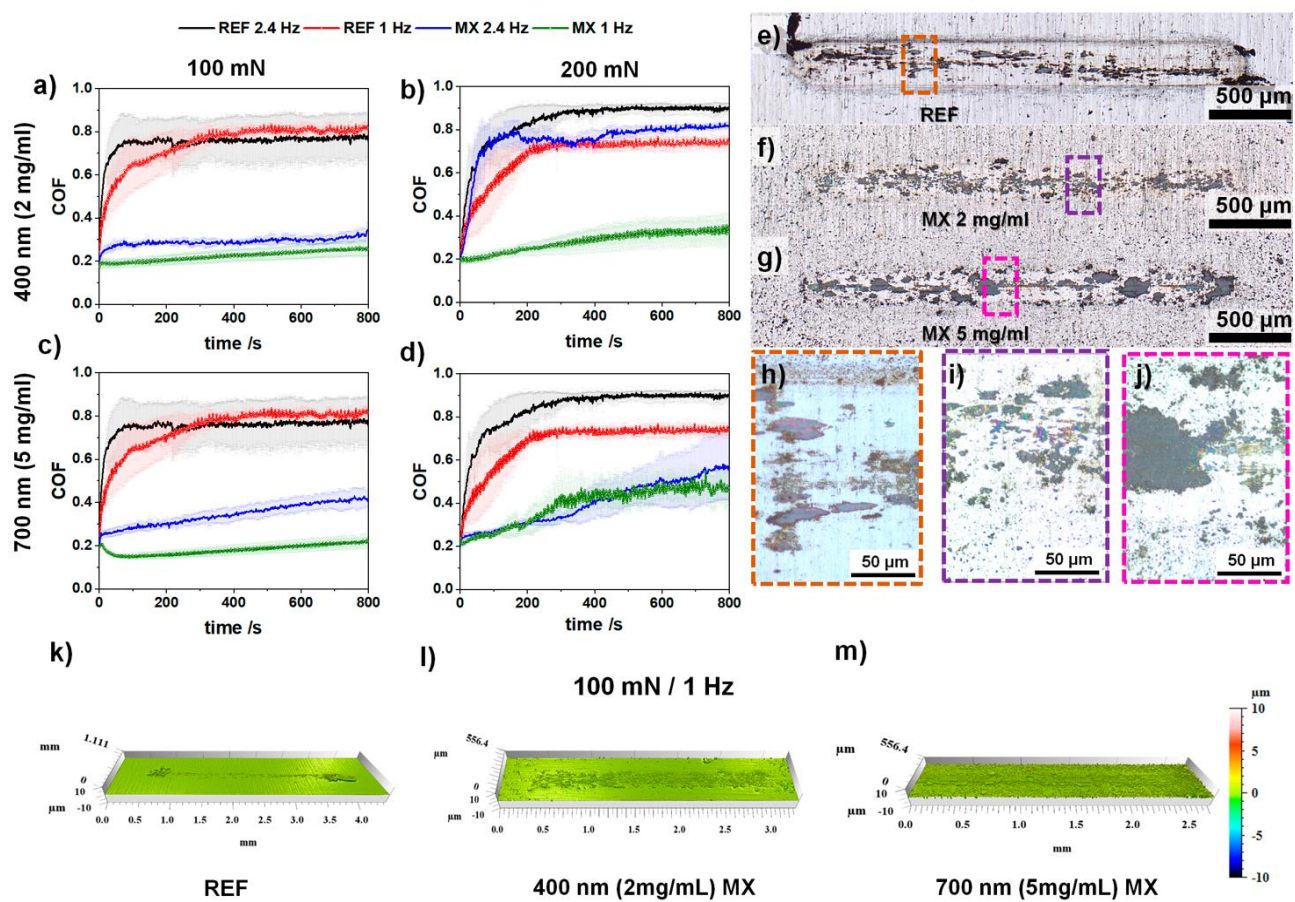

**Figure 2**. (**a-d**) Temporal evolution of the coefficient of friction of multi-layer $Ti_3C_2T_x$ solid lubricant coatings with an initial thickness of (**a** and **b**) 400 nm (dispersion concentration of 2 mg/ml) and (**c** and **d**) 700 nm (5 mg/ml) measured at (**a** and **c**) 100 and (**b** and **d**) 200 mN. All tribological experiments were conducted at two different frequencies (1 and 2.4 Hz), which is indicated by the given color code. (**e-g**) Light microscopic images and (**h-j**) corresponding inserts with enhanced magnification of the corresponding wear tracks of (**e** and **h**) the uncoated reference as well as (**f, g, i** and **j**) the $Ti_3C_2T_x$ solid lubricant coatings with an initial thickness of (**f** and **i**) 400 nm (2 mg/ml) and (**g** and **j**) 700 nm (5 mg/ml). (**k, l,** and **m**) 3D white light interferometric images of the wear tracks of (**k**) the uncoated reference and (**l** and **m**) $Ti_3C_2T_x$ coatings with an initial thickness of (**l**) 400 nm (2 mg/ml) and (**m**) 700 nm (5 mg/ml). Please note that multi-layer $Ti_3C_2T_x$ nano-sheets have been abbreviated by "MX" in this figure.

Next, the friction (**Figure 2**) and wear performance of all samples (**Figure S3**) dependent on the normal load (100 and 200 mN) and sliding frequency (1 and 2.4 Hz) were studied. We first tested the steel substrates as our reference (**Figure 2a-d**), which show the typical running-in known for steel-steel contacts under dry sliding conditions irrespective of the testing conditions[36, 37]. The initial coefficient of friction (COF) was about 0.25 before increasing almost instantaneously to ~ 0.8 - 0.9 and finally reached steady-state conditions after a sliding time of about 200 - 300 s. In case of 100 mN, the steady-state COF obtained for different frequencies is almost identical, reaching values of about 0.7. Minor differences in the resulting steady-state COF can be seen for a normal load of 200 mN reaching values of 0.7 and 0.8 for frequencies of 1 and 2.4 Hz, respectively. In this regard, an increased normal load may induce more plastic deformation, higher temperatures, and more severe oxidation, thus leading to an increased COF [38]. Additionally, the sliding time to reach steady-state conditions is slightly prolonged for the lower frequency of 1 Hz, which may be connected to the reduced frictional energy induced into the tribological contact.[38]

The COF for multi-layer $Ti_3C_2T_x$ coatings (**Figure 2a-d**) started with an initial value of about 0.2, irrespective of the considered thickness (concentration) and normal load. In most cases, the $Ti_3C_2T_x$ coatings show an improved tribological performance with a reduced COF and more stable evolution, which holds especially true for a sliding frequency of 1 Hz. There are some exceptions for the measured COF of $Ti_3C_2T_x$, such as for 200 mN and a thickness of 700 nm (concentration of 5 mg/ml), which will be discussed below.

In the case of 100 mN and a thickness of 400 nm (2 mg/ml), only minor differences in the absolute values and temporal evolution of the COF for both tested frequencies can be seen (**Figure 2a**). For the same normal load of 100 mN, differences became more pronounced when increasing the coating thickness to 700 nm (5 mg/ml) because of increasing the amount of lubricious material (compare green with blue in **Figure 2c**). While the frictional evolution for 2.4 Hz shows a continuously rising trend without reaching steady-state conditions, the experimental trend for 1 Hz is rather different. In the first sliding cycles, the COF decreases before leveling off and maintaining stable values of around 0.2 over the entire measuring time. The initial decrease may be connected with the partial reorientation of nano-sheets parallel to the surfaces as observed and verified for other 2D nanomaterials such as $MoS_2$ [39, 40].

For a normal load of 200 mN (**Figure 2b** and **d**), the experimental trends notably change as compared to the 100 mN normal load. Experiments performed with 2.4 Hz demonstrate an instantaneous increase in the COF thus reaching steady-state values of the uncoated reference (**Figure 2b**). In contrast, a beneficial evolution with only a marginal COF increase reaching steady-state values of about 0.3 can be observed for experiments performed with 1 Hz., By a further increase of the coating thickness to 700 nm (MXene concentration of 5 mg/ml), the frictional performance for a frequency of 2.4 Hz was improved, which can be a result of the increased amount of lubricious material in the contact. In case of 1 Hz, a similar frictional evolution as compared to the lower thickness (concentration of 2 mg/ml) can be seen for the 200 mN load (**Figure 2b and d**). These experimental tendencies demonstrate that the initial MXene coating thickness and adjusted sliding frequency affect the tribological behavior, while the best frictional performance was found for the thinner MXene coating and lower frequency.

After rubbing, the wear features were studied by light microscopy in DIC (**Figure 2e-j**) and WLI (**Figure 2k** and **l**). The uncoated reference shows clear abrasion marks with deposited debris at the reversal points and across the entire perimeter (**Figure 2h** and **k**). Moreover, black areas inside the wear track point towards oxidative wear (**Figure 2h**). In contrast, the multi-layer $Ti_3C_2T_x$ coatings do not show pronounced abrasion or deposited debris in or around the wear track (**Figure 2i, j** and **l**). Using DIC (**Figure 2i** and **j**), patchy-like structures across the entire wear-track show a blueish or greenish color, which is indicative of tribo-chemical reactions due to the phase sensibility of DIC. For 1 Hz, these structures tend to increase in size and cover more areas of the wear track when increasing the concentration to 5 mg/ml (**Figure 2j**). To elucidate the involved processes and mechanisms, the tribological interfaces were characterized by TEM.

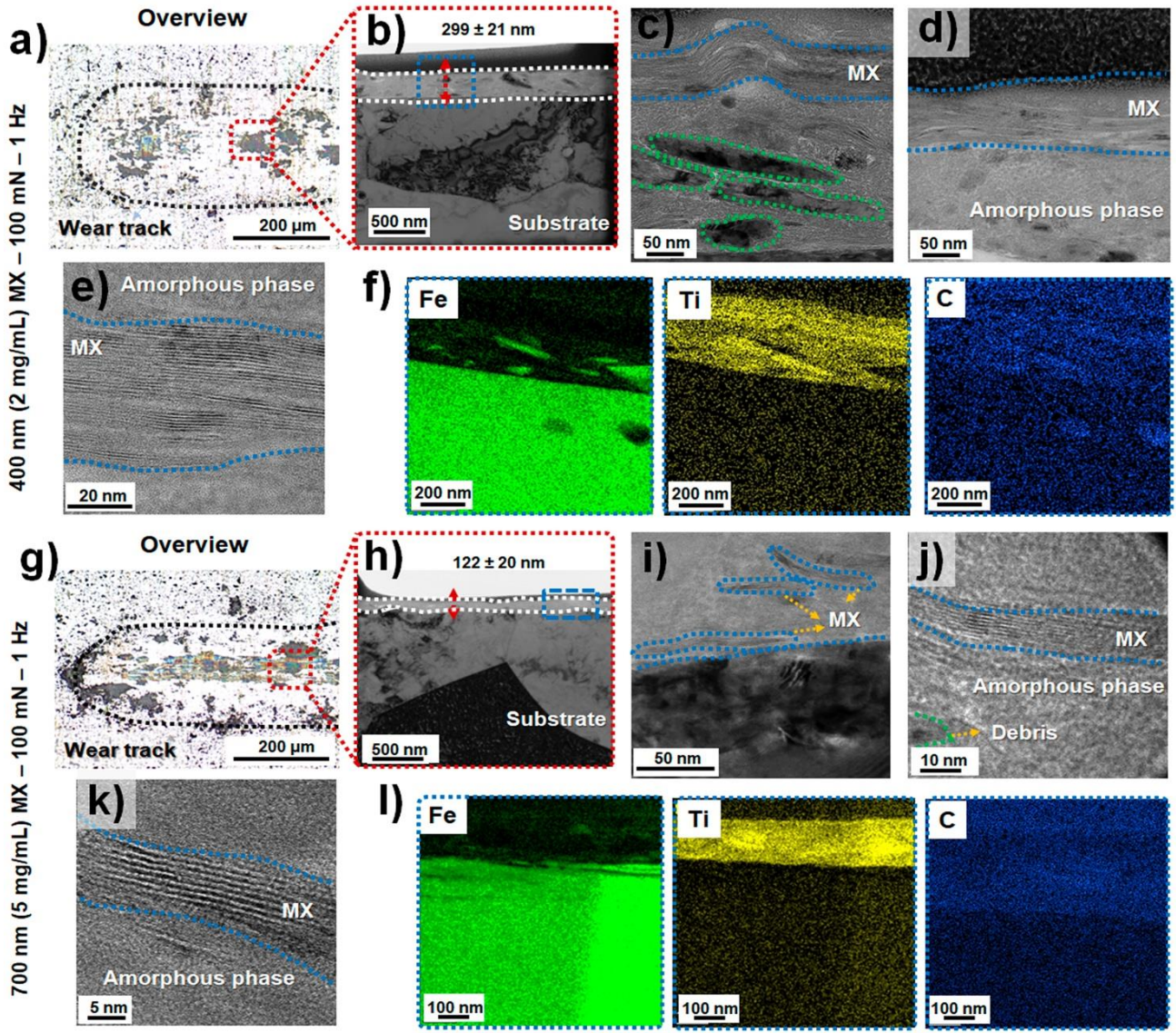

**Figure 3**. (**a** and **g**) Light-microscopic images acquired in DIC of the resulting wear tracks for multi-layer $Ti_3C_2T_x$ coatings with an initial thickness of (**a**) 400 nm (dispersion concentration of 2 mg/ml) and (**g**) 700 nm (5 mg/ml) measured for a normal load of 100 mN and a frequency of 1 Hz. These DIC images helped to identify the respective positions to prepare the TEM lamellas for the interface characterization by FIB. (**b-e** and **h-k**) TEM micrographs of the tribological interfaces showing the formed tribo-layers for the $Ti_3C_2T_x$ coatings with an initial thickness of (**b-e**) 400 nm (2 mg/ml) and (**h-k**) 700 nm (5 mg/ml). The formed tribo-layer is marked in (**b** and **h**) by white dashed lines, while remaining layered 2D MXene structures are highlighted with blue dashed lines in (**c-d**) and (**i-k**), respectively. Intermixed wear debris (**c** and **j**) originating from the initial steel substrate is marked with green dashed lines. (**f** and **l**) TEM-EDX maps showing the distribution of Fe, Ti and C in the tribo-layer for coatings with an initial thickness of (**f**) 400 nm (2 mg/ml) and (**l**) 700 nm (5 mg/ml), respectively. The respective locations, in which the TEM-EDX maps have been acquired, are marked in (**b** and **h**) by blue rectangles. Please note that multi-layer $Ti_3C_2T_x$ nano-sheets have been abbreviated by "MX" in this figure.

The best tribological performance with the lowest and most stable COF was observed for multi-layer $Ti_3C_2T_x$ coatings (2 and 5 mg/ml, which correspond to thicknesses of 400 and 700 nm, respectively) at 100 mN and 1 Hz (**Figure 2a** and **c**). The wear tracks imaged by light microscopy in DIC (**Figure**

**3a** and **g**) demonstrate patchy-like features, which increase with increasing MXene concentration. Together with the observed color change (blueish/greenish), these aspects are indicative of the formation of a tribo-layer. Moreover, these positions match with increased amounts of titanium originating from the initial $Ti_3C_2T_x$ nano-sheets as verified by EDX [21]. Since the high-resolution interface characterization relies on the fabrication of TEM specimens, which is time-consuming and highly localized, this simple color change due to the phase sensibility of DIC helps to unambiguously identify the positions of interest with the highest probability of finding remaining $Ti_3C_2T_x$ in the tribo-layer.

Low-magnification TEM micrographs of the FIBed cross-section area in the wear track verify the formation of a tribo-layer with thicknesses of about 299 (**Figure 3b**) and 120 nm (**Figure 3h**) for initial MXene concentrations of 2 and 5 mg/ml, respectively. A comparison of the as-deposited $Ti_3C_2T_x$ (**Figure 1h** and **i**) with the newly formed interface makes evident that the tribo-layer is well adhered to the substrate without the appearance of any defects, cracks, or voids. We believe that this implies the formation of a strong tribo-layer/substrate interface (**Figure 3b** and **h** as well as **Figure S4a**). A more detailed characterization of the existing features and structures in the tribo-layers with higher magnification (**Figure 3c-e** and **i-j** as well as **Figure S4b-d** and **Figure S5**) reveals differences depending on the initial coating thickness.

For the thinner (400-nm-thick) coatings (**Figure 3c-e** and **Figure S4b-d**), the tribo-layer still contains multi-layer $Ti_3C_2T_x$ with a reduced multi-layer thickness compared to their initial state (**Figure 1b**). For the thicker (700-nm-thick) coatings (**Figure 3i-k** and **Figure S5**), however, only few-layer $Ti_3C_2T_x$ can be found in the formed tribo-layer. This observation points towards tribo-induced exfoliation processes that is dependent on the initial coating thickness, which appears to impact the MXene features in the final tribo-layer. We note that irrespective of detected multi- or few-layer nano-sheets, they are intermixed with wear debris originating from the steel substrate as well as amorphous and nanocrystalline oxidic structures (**Figure 2d** and **j, Figure S4b-d,** and **Figure S5**)).

Moreover, it becomes evident that the detected MXenes are thinner than the original multi-layer MXene particles (**Figure 1h** and **i**) and they are parallelly aligned with respect to the substrate (**Figure 3d** and **e** as well as **Figure S4b-d**). This implies that they might have undergone a stress-induced exfoliation and reorientation thus aligning their basal plane towards the direction of the acting shear stress. In this manner, it becomes clear that the MXene sheets orient themselves along the direction of their lowest shear strength, which results in MXenes' basal plane orienting parallel to the acting shear force on the tribo-layer. As a result of this observation, we believe that one particularly impressive point is the observed bending flexibility of the multi-layer MXenes in the tribo-layer [41] (**Figure 3c**), which appears to help the alignment of the nano-sheets with the underlying structure and to accommodate any underlying roughness/topography/feature. The TEM-EDX maps shown in **Figure 3f** and **l** show a rather homogeneous distribution of titanium and carbon throughout the entire tribo-layer, which stem from the initial MXenes deposited onto the steel substrates. Interestingly, iron is also detected in the MXene region (**Figure 3f**), which suggests the parallel-oriented MXene area in this tribo-layer is formed during the wear process. This implies that the observed MXene features are still rather homogeneously distributed in the tribo-layer covering the wear track, which likely contributes towards the improved frictional performance.

Regarding the observed thickness-dependent exfoliation processes, similar features can be seen in the formed tribo-layers when tested at a frequency of 2.4 Hz (**Figure S6**). The tribo-layer of the thinner coating (**Figure S6a-f**) reveals a homogeneous thickness (**Figure S6b**), the existence of few-layer MXenes (**Figure S6c**) as well as a homogenous Ti distribution across the entire tribo-layer (**Figure S6f**). The tribo-layer of the thicker coating demonstrates a more inhomogeneous thickness (**Figure S6h**) and a stochastic Ti distribution across the tribo-layer (**Figure S6l**), which is not indicative of any existence of remaining MXene sheets in the tribo-layer. Compared with the thinner coating tested at 1 Hz (**Figure 3a-f**), which revealed multi-layer in the tribo-layer, this is indicative that the involved exfoliation processes do not only depend on the initial coating thickness but also the adjusted sliding velocity.

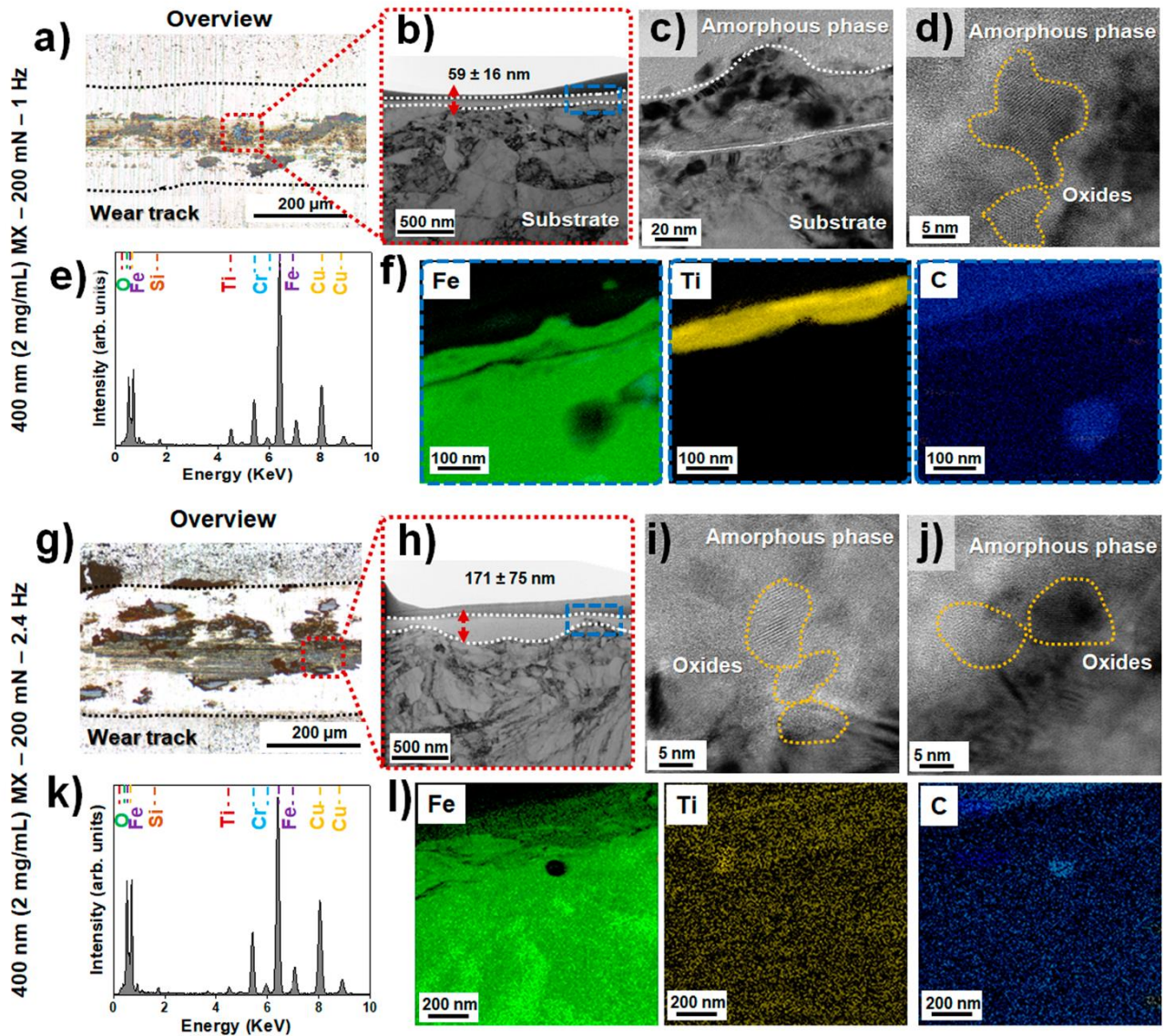

**Figure 4**. (**a** and **g**) Light-microscopic images acquired in DI contrast of the resulting wear tracks for multi-layer $Ti_3C_2T_x$ coatings with an initial thickness of 400 nm (dispersion concentration of 2 mg/ml) measured for a normal load of 200 mN and sliding frequencies of (**a**) 1 and (**g**) 2.4 Hz, respectively. These images helped to identify the respective positions to prepare the TEM lamellas for the interface characterization. (**b-d** and **h-j**) TEM micrographs of the resulting tribological interfaces showing the formed tribo-layers for the $Ti_3C_2T_x$ coatings (400 nm, 2 mg/ml) for a normal load of 200 mN and sliding frequencies of (**b-d**) 1 and (**h-j**) 2.4 Hz, respectively. The formed tribo-layer is marked in (**b** and **h**) by white dashed lines, while oxidic structures are highlighted with yellow dashed lines in (**d**) and (**i** and **j**), respectively. (**e** and **k**) EDX-spectra with (**f** and **l**) the corresponding element mapping for Fe, Ti and C for wear tracks of multi-layer $Ti_3C_2T_x$ coatings (400 nm, 2 mg/ml) measured for a normal load of 200 mN and sliding frequencies of (**e** and **f**) 1 and (**k** and **l**) 2.4 Hz. The respective locations, in which the TEM-EDX maps have been acquired, are marked in (**b** and **h**) by blue rectangles.

After having explored the interfacial processes for multi-layer $Ti_3C_2T_x$ coatings at 100 mN and 1 Hz, the tribological interface was characterized for the thinner $Ti_3C_2T_x$ coating (400-nm-thick made from 2 mg/ml) tested at 200 mN for both sliding frequencies (**Figure 4**). At the higher load of 200 mN, the coatings show pronounced differences depending on the adjusted frequency. As discussed earlier (**Figure 2b**), a stable and low COF for 1 Hz was observed, whereas the COF instantaneously increases for 2.4 Hz thus reaching the steady-state values of the reference of about 0.73 after 200 s. For the thin $Ti_3C_2T_x$ coatings (2 mg/ml) tested at 1 Hz, it becomes evident that the resulting wear tracks show the same blueish/greenish color change in DIC (**Figure 4a**). The low-magnification TEM images (**Figure 4b**) show a tribo-layer with a reduced but still relatively uniform thickness, while high-resolution TEM images (**Figure 4c** and **d**) reveal the existence of amorphous (short-range order as additionally demonstrated in **Figure S7**) and nanocrystalline oxidic species. In the acquired high-resolution micrographs, no layered structures were observed. However, the detailed EDX analysis clearly identifies a pronounced Ti signal (**Figure 4e**). The respective TEM-EDX maps depicted in **Figure 4f** shows a strong Ti signal throughout the entire tribo-layer, which points towards the existence of MXenes in the tribo-layer.

In contrast, the interfacial characterization of the wear tracks for the thin $Ti_3C_2T_x$ coatings (400-nm-thick, 2 mg/ml) tested at 200 mN and 2.4 Hz shows a very different picture. In the light-microscopic DIC image (**Figure 4g**), blackish, patchy-like features without any color change are detectable. The low-magnification TEM image (**Figure 4h**) demonstrates the formation of a tribo-layer with a nonuniform thickness with variations of ± 75 nm in a ~170-nm-thick layer. High-resolution TEM micrographs (**Figure 4i** and **j**) verify the existence of amorphous and oxidic structures. When checking the acquired EDX spectrum (**Figure 4k**), a notably reduced Ti signal becomes evident. This aligns well with the presented elemental mapping (**Figure 4l**), which only shows a stochastic distribution of Ti throughout the entire wear-tracks. Therefore, we anticipate that the increased normal load and frequency removed MXenes from the tribological interface, thus depositing them at the reversal points. This is consistent with the stochastic Ti signal in the wear track, which is not indicative of any pronounced Ti accumulation and, therefore, the existence of remaining MXenes in the wear track. Consequently, the tribo-layer is mainly constituted of amorphous and nanocrystalline oxidic species, which implies that oxidative wear is the governing wear mechanism. This aligns well with the observed light-microscopic features and the instantaneous increase of the COF.

Summarizing, the conducted tribological experiments on multi-layer $Ti_3C_2T_x$ coatings unambiguously verified that tribo-chemical and exfoliation processes govern their underlying tribo-layer formation and, consequently, solid lubrication ability. To explain these experimental trends, different contributions need to be taken into consideration. Compared to the initial coating quality, which verified the deposition of individual, disconnected accordion-like multi-layer particles with a weak substrate-coating interface (**Figure 1h, i**), the high-resolution interface characterization after the tests demonstrated that the nano-sheets experienced tribo-induced exfoliation processes, which reduced their thickness (z-dimension) thus resulting in multi- or few-layers being present in the formed tribo-layers. Additionally, our results indicated that the nano-sheets underwent a reorientation of their basal plane during rubbing (shearing). This helped to match their orientation with the lowest shear strength with the acting external shear stress. Only for low normal load and low-velocity conditions combined with thinner coatings, homogeneous, uniform, densified, well-adhering tribo-layers containing highly aligned and exfoliated nano-sheets intermixed with amorphous and nanocrystalline oxide species were formed (**Figure S7**), which induced low friction and an excellent wear-resistance.

When increasing either frequency (2.4 Hz), normal load (200 mN) or coating thickness (~700-nm-thick coatings, 5 mg/ml), no signs of either titanium or only highly exfoliated few-layer MXenes were found in the wear track. This implies that higher frequencies and normal loads represent more severe testing conditions, which may move the initially deposited, poorly adhered nano-sheets towards the reversal points of the wear tracks. For thicker coatings, the existence of few-layer MXenes points towards an increased exfoliation probability, which implies a more severe degradation. This reduces the amount of lubricious MXenes in the tribo-layer thus detrimentally affecting the resulting frictional performance.

To evaluate the importance of MXene multi-layer particles in friction, we delaminated the multi-layer MXene particle to few-layer $Ti_3C_2T_x$ by further processing (see Methods section) and prepared another set of coatings with the few-layer $Ti_3C_2T_x$ solutions. We kept the coating thicknesses and tribological testing conditions the same as for the multi-layer particle coatings (**Figure S7**). Compared to multi-layer coatings (**Figure 2**), few-layer $Ti_3C_2T_x$ coatings show a rather downgraded frictional behavior. Irrespective of the tribological testing conditions and coating thickness, most of the coatings made from the few-layer MXene show an increasing tendency over time, which does not reflect any beneficial friction behavior and/or wear resistance. For the thinner coatings (**Figure S7a** and **b**), an improved frictional evolution can be seen for the lower sliding velocity (1 Hz), which is consistent with the trends observed for the coatings made from the multi-layer MXene particles. Thicker coatings (**Figure S7c**) do not show any pronounced velocity-dependency. The downgraded friction and wear performance of the coatings made from the delaminated few-layer MXene can be connected to the lower thermal stability of few-layer $Ti_3C_2T_x$ nano-sheets, which makes them more prone to degradation and oxidation. When assuming similar thicknesses, few-layer coatings have more weakly connected interfaces, which increases the likelihood of being shifted to the reversal points without further need for exfoliation. Consequently, the initial z-dimension (layer thickness) together with the reduced thermal stability of few-layer $Ti_3C_2T_x$ favor their full degradation/removal thus making it highly unlikely that any lubricious nano-sheets remain in the tribo-layers. This aligns well with the downgraded tribological performance observed for few-layer $Ti_3C_2T_x$ coatings.

To further shed light on the observed thickness- and velocity dependency, we utilized MD simulations of sliding MXene films with different thicknesses using the interatomic interactions by a ReaxFF potential as previously designed for titanium carbide MXenes [42]. **Figure 5a** schematically depicts an exemplary model system of a titanium carbide $Ti_2CO_2$ MXene film with four layers ($N = 4$). We used $Ti_2CT_x$ instead of $Ti_3C_2T_x$ as we only investigated the interaction between the MXene flakes (inter-flake interactions). We have previously shown that the MXene flake thickness ($Ti_2CT_x$ to $Ti_4C_3T_x$) does not affect the work of separation of MXenes [43]. To conduct our simulations, these layers are sandwiched between a fixed bottom layer, mimicking anchoring onto an adhesive substrate, and a top layer that mimics a slider, e.g., a layer attached to a tribological counter-body. We studied the effect of coating thickness by varying $N$ between 1 and 19. Under an applied, constant pressure of 5 GPa, the top-most Ti atoms of the slider were moved at constant velocities of 10 and 50 m/s to demonstrate a potential effect of the sliding velocity. In a previous study based on first principles calculations, we showed that an acting normal stress tends to increase the corrugation of the potential energy surface (PES) experienced by the sliding layers, which results in higher frictional forces thus being in good agreement with experimental results [43].

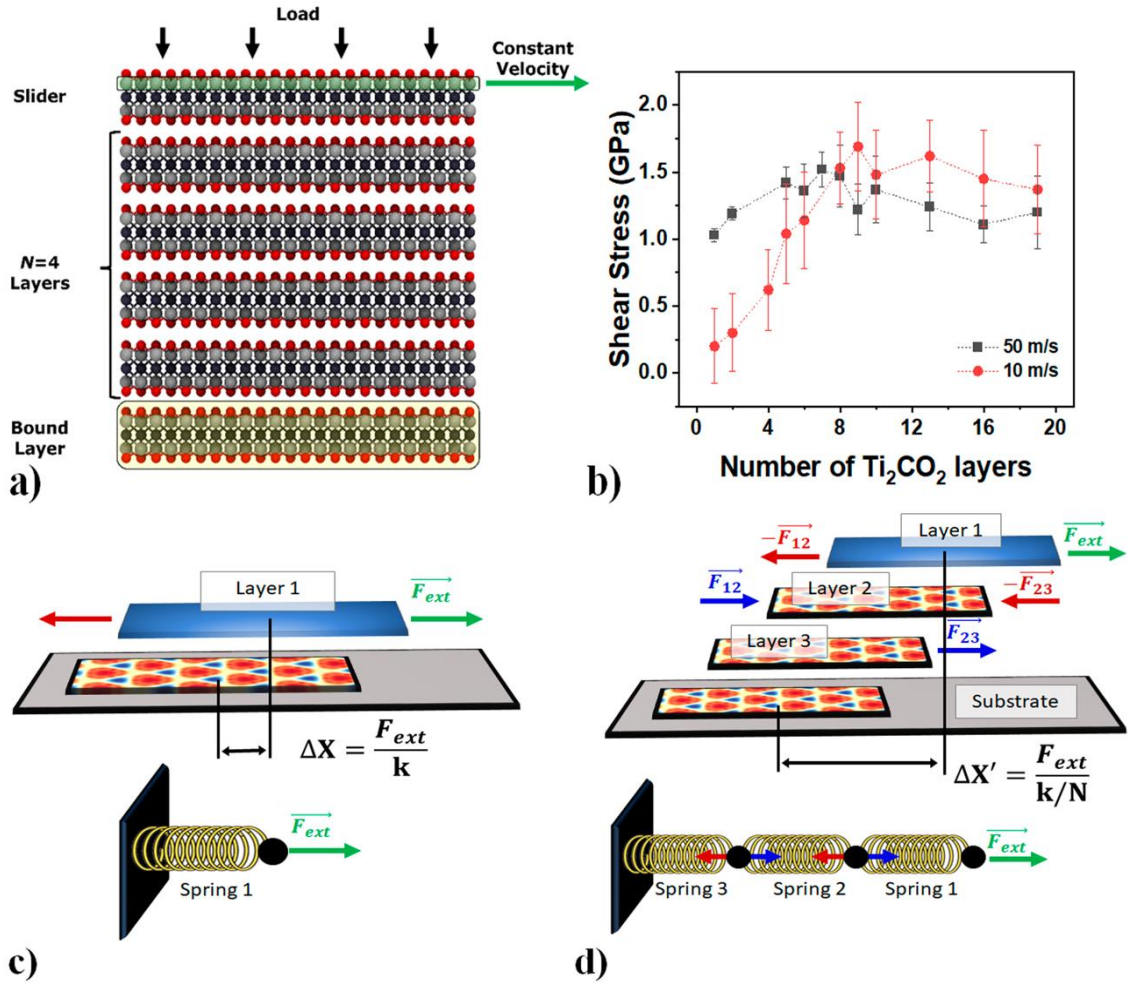

**Figure 5.** (**a**) Schematic representation of the Ti$_2$CO$_2$ system having 4 layers. The bottom layer ("bound layer") highlighted in yellow color was fixed. A normal pressure of 5 GPa (indicated by "load") was applied to the topmost Ti atoms of the top layer ("Slider"), while the top layer was moved at a constant velocity along the x direction. Color code: Ti, C and O atoms are depicted in grey, black, and red, respectively. (**b**) Shear stress (GPa) experienced by the slider during 270 ps of simulation as a function of the film thickness for two different sliding velocities. Red dots and black squares refer to drag velocities of 10 and 50 m/s, respectively. (**c**) Schematic representation of *N* stacked layers interacting with each other through a potential energy surface. In this regard, layer 1 is pulled out from the stack, while layer *N* is anchored to the substrate. The forces on the layers are similar to elastic forces on a series of springs as illustrated in (**d**).

After reaching thermal equilibrium, a simulation of 270 ps was carried out for all considered MXene systems. During this time interval, the slider was laterally displaced by 27 Å (135 Å) in the low (high) velocity regime. The lateral motion of the slider was partially transferred to the underlying MXene layers, which were dragged along the sliding direction, resulting in energy dissipation. The mean values of the restoring shear stress on the slider center of mass are depicted in **Figure 5b** as a function of the number of Ti$_2$CO$_2$ layers and sliding velocity. In the high velocity regime (orange dots), the shear stress is almost independent of the film thickness. In contrast, in the low velocity regime (green dots), the shear stress observed for thinner films turned out to be significantly lower than for thicker films. These differences can be observed on the left-hand side of the graph, which corresponds to thinner coatings ($N < 6$), while, for $N > 6$, the resistance force on the slider becomes almost independent of the sliding velocity.

The MD results qualitatively agree with the observed experimental trends for few-layer (**Figure S7**) and multi-layer MXene coatings (**Figure 2**) tested at different sliding frequencies. Indeed, the COF of the thicker MXene films is almost independent of the sliding velocity/frequency, while thinner films show significantly lower friction at lower sliding velocities. This goes hand in hand with previous MD simulations of multi-layer graphene films performed by the authors, which revealed that stacked layers behave as springs added in series.[44] We anticipate that this is a general property of 2D materials that also holds true for MXenes.

As depicted in **Figure 5c and d**, the interaction between two adjacent layers is ruled by the PES that describes the variation of the interlayer energy as a function of the relative lateral position of each layer. When the upper layer ("layer 1") is displaced from its initial position, corresponding to a PES minimum, it experiences a restoring force -$F_{12}$. At the same time, it imposes an opposite force on "layer 2", $F_{12}$. Since the multi-layer structure is homogeneous, the PES is common to each bilayer, the magnitude of these relative forces is, thus, the same for all N layers ($F_{12} = F_{23}$, etc.). When approximating the PES in close proximity of the minimum with a harmonic potential, its derivative corresponds to an elastic force, which can be analogously represented by springs. A film composed of $N$ layers can be treated as springs connected in series having a spring constant, which decreases with film thickness. Consequently, the higher the number of layers, the lower the resistance to exfoliation of the topmost layer, which aligns well with the presented experimental data.

## 4. Conclusions

This study aimed at exploring the solid lubrication performance of MXene coatings dependent on the initial MXene state (multi-layer versus delaminated, few-layer $Ti_3C_2T_x$), coating thickness (400 versus 700 nm), applied normal load (100 versus 200 mN) and sliding frequency (1 versus 2.4 Hz). The best tribological performance with the lowest and most stable COF was observed for multi-layer $Ti_3C_2T_x$ coatings tested at the lower normal load (100 mN) and lower frequency (1Hz). For these conditions, the high-resolution interface TEM characterization revealed the formation of a uniform, well-adherent, densified tribo-layer containing different oxidic species intermixed with exfoliated, multi-layer $Ti_3C_2T_x$ with an orientation parallel to the underlying substrate.

When using delaminated, few-layer MXenes to create the initial coatings or increasing either frequency (2.4 Hz), normal load (200 mN) or coating thickness (~700-nm-thick coatings, 5 mg/ml), a less beneficial or even detrimental friction evolution with an instantaneous increase reaching COF values of the uncoated steel references is observed. The high-resolution interface characterization revealed the presence of an inhomogeneous tribo-layer mainly constituted of oxides and partially highly exfoliated few-layer MXenes. In this regard, higher frequencies and normal loads represent more severe testing conditions thus shifting more lubricious material towards the reversal points of the wear tracks. For thicker coatings, the existence of few-layer MXenes points towards an increased exfoliation probability, which implies a more severe degradation. Coatings made of few-layer MXenes tend to have more internal interfaces, which can accelerate the removal/shifting of MXenes to the reversal points or increase their exfoliation probability.

Therefore, we conclude that MXenes' solid lubrication ability depends on the involved tribo-chemical processes, but also tribo-induced exfoliation as confirmed by the high-resolution interface characterization together with reactive MD modelling. In this regard, MD simulations clearly demonstrated that an increasing number of layers leads to reduced resistance to exfoliate the topmost layer. Consequently, to induce a low friction and wear-resistant performance, coatings made of multi-layer Ti$_3$C$_2$T$_x$ MXenes must be optimized regarding the tribological testing conditions (interplay of thickness, load, and testing frequency) to ensure the formation of a uniform, well-adherent tribo-layer, which still contains aligned, multi-layer MXenes thus limiting tribo-induced exfoliation.

**Conflict of Interest**

The authors declare that they have no known competing financial interests or personal relationships that could have appeared to influence the work reported in this manuscript.

**Funding**

A. R. gratefully acknowledges the financial support given by ANID-CONICYT within the projects Fondecyt Regular 1220331 and Fondequip EQM190057. J. Y. A.-H., D. Z. and A. R. acknowledge the final support given by ANID-Chile within the project Fondecyt Postdoctorado 3210052 y 3220165. B. W. acknowledges the financial support provided by the Alexander von Humboldt Foundation. M.C.R. and E.M. gratefully acknowledge the project "Advancing Solid Interface and Lubricants by First Principles Material Design (SLIDE)" funded by the European Research Council (ERC) under the European Union's Horizon 2020 research and innovation program (Grant agreement No. 865633).

**Authors contributions**

A. R. came up with the initial idea and took the lead of the project. J. M. H., D. Z., J. Y. A.-H., and A. R. conducted the tribological experiments and analyzed the results. B. W. conducted the high-resolution interface characterization by TEM. E. M., P. R. and M. C. R. performed the atomistic and reactive molecular dynamic simulations. B. C. W. and B. A. contributed towards the analysis of the as-synthesized coatings as well as the overall discussion of the underlying MXene-related mechanisms. All authors contributed towards the discussion of the obtained results. All authors have worked, proofread, and approved on the final draft of the manuscript.

**Data Availability Statement**

Data are available upon request with the corresponding author.